\documentclass[twocolumn,showpacs,preprintnumbers,amsmath,amssymb]{revtex4}
\usepackage{graphicx}
\usepackage{dcolumn}
\usepackage{bm}
\begin{document}
\preprint{}
\title{A New Interpretation on Quantum Mechanics}
\author{Guang-jiong Ni}
\email{gj_ni@Yahoo.com}
\affiliation{Department of Physics, Fudan University, Shanghai, 200433, China\\
Department of Physics, Portland State University, Portland, OR 97207, USA}
\begin{abstract}
Based on new experiments about the ``macroscopic Schrodinger's cat state" etc., a
self-consistent interpretation on quantum mechanics is presented from the new point of
view combining physics, philosophy and mathematics together.
\end{abstract}
 \maketitle
\section{\label{sec:1}Introduction}
Quantum mechanics was established in 1925. However, in 1964, in his lecture at Cornell
University, the famous physicist Feynman said: ``There was a time when the newspapers
said that only twelve men understood the theory of relativity. I do not believe there
ever was such a time. \dots On the other hand, I think I can safely say that nobody
understands quantum mechanics. \dots Do not keep saying to yourself, if you can possibly
avoid it, `But how can it be like that?' because you will get `down the drain', into a
blind alley from which nobody has yet escaped. Nobody knows how it can be like that."
[1][2].

Why quantum mechanics is so difficult to understand? Let's look at a conceptual
experiment proposed by Schrodinger in 1935 (see e.g.[2] [3]): A cat is confined in a
closed box and staying in a quantum (coherent) state described by the wavefunction:
\begin{equation}\psi(x,t)=\psi_0(x,t)+\psi_1(x,t)\label{eq:1},\end{equation}
Here $\psi_0$ and $\psi_1$ denote the ``dead cat" and ``alive cat" states respectively
with $x$ and $t$ being the space and time coordinates. Once when we open the box to
observe, a process of so-called ``the collapsing of wave-packet" occurs suddenly. The
probability of discovering the cat being dead (or alive) would be proportional to
$|\psi_o|^2$ ( or $|\psi_1|^2$ ). Such kind of way of saying sounds absurd and so is
named a paradox, aiming at rendering the contradiction contained in the theory more
acute. It might be helpful to clarifying what's wrong in the basic concept? The ``cat
paradox" raised by Schrodinger at least posed the following four questions in front of
all physicists in a very acute way:

1. Is the Born statistical interpretation of quantum mechanics correct?

2. Is this probability interpretation valid for single system (a pure state, e. g. an
electron or a cat)? Or we need an average over many systems under the same conditions
(so-called the ensemble).

3. Is it suitable to describe a macroscopic living-being like a cat by the wavefunction
of quantum mechanics?

4. If the cat is shrinking into a small one such that it can be described by the
wavefunction~(\ref{eq:1}), is it suitable to talk about $\psi_0$ and $\psi_1$ as two dead
and alive cat states respectively?

\section{\label{sec:2}Experiments on Schrodinger's Cat State}
After research for more than 70 years, the Born statistical interpretation has been
verified by numerous scattering and decay experiments and it is valid even for single
system. However, some subtlety in the probability interpretation was often overlooked
(see below). The answer to the third question could be conjectured quite early: Only an
object within the ``quantum coherent length" can be described by the wavefunction , it is
impossible for a large living-being like a real cat. The theoretical research in 1980s
revealed that the principal factor limiting the coherent length of a quantum system is
the ``decoherence" effect induced by the coupling between the system and its environment.
In recent years, the microscopic (atomic scale) and mesoscopic (in nm scale)
Schrodinger's cat states have been realized successively in various experiments.

In an experiment published in January 2000[4], Myatt et al. manufactured delicately the
environment of single $^9Be^+$ ion (the microscopic Schrodinger's cat state) captured in
an ion trap and observed the process of decoherence, i.e., the process of how the ``cat"
is killed during its interaction with environment. In July 2000, Friedman et al. prepared
a ring of superconducting-quantum-interference-device (SQUID) under the magnetic field at
low temperature[5]. Then the superconducting state inside the ring can be described by
Eq.~(\ref{eq:1}). It was stressed that $\psi_0$ and $\psi_1$ denote two states with
clockwise and counterclokewise currents respectively and they differ in current 2-3
m$\mu$A, corresponding to the motion of $10^9$ electron (Cooper) pairs in opposite
directions (Here the coordinate $x$ in Eq.~(\ref{eq:1}) represents the magnetic flux
inside the ring). Since $10^9$ is a large number, the above experiment for the first time
realized the thought experiment of Schrodinger's in a truly macroscopic scale with
$\psi_0$ and $\psi_1$ as the simulation of dead-and-alive cat.

Although $\psi_0$ and $\psi_1$ are two distinct states and if $|\psi_1|>|\psi_0|$, there
is only counter clockwise superconducting current superficially, the microwave absorption
experiment [5] does verify the existence of $\psi_0$ as shown by the absorption
probability which is proportional to $|\psi_0|^2$. On the contrary, if
$|\psi_0|>|\psi_1|$, there is only clockwise current superficially, the existence of
$\psi_1$ is also verified by experiments. When $|\psi_0|=|\psi_1|$, there is no current
superficially, but the experimental analysis does show that two directions of current
exhibit their existence with equal probabilities. All experimental results are in
conformity with the calculation of quantum mechanics.

In the same issue of Nature journal, a theorist Blatter wrote a paper titled
``Schrodinger's cat is now fat" to explain the experiment [6]. And some articles were
published on the famous media like New York Times to introduce the new experiment. In
some caricature an alive cat with dead cat as its overlapping shadow was even larger than
a man and so attracted strong interests in the public. However, ``there is dead-cat
inside the alive-cat" or ``there is alive-cat inside the dead-cat", do you believe in it?

\section{\label{sec:3}Where is the Problem?}
Just like some problem in other field, once the contradiction becomes more acute, we are
more near to the solution for the problem. Now the cat becomes bigger and bigger whereas
quantum mechanics works better and better. To our understanding, new experiments are just
further verifying that the cat is nothing but an ``illusion" existing only in our brain
which in turn proves that our interpretation on quantum mechanics was incorrect in the
sense that the ``quantum state" and ``wavefunction' were understood too ``materialized"
in the past. Hence in our opinion, the "cat paradox" is now basically over [7].

Let's consider a concrete example. A big company is controlled by two stockholders A and
B. While holding stocks up to 51\%, A has the management idea to run the enterprise
``eastward". On the contrary, B who holds stocks up to 49\% claims that the enterprise
should go ``westward". Then the board of directors passes a resolution to follow A's
opinion that the whole enterprise should carry out the management policy running
``eastward". Despite 49\% of all staff members being reluctant to do so, all members are
uniform in their paces. Alternatively, if due to some abrupt change, the stocks hold by B
rise to 51\% while that by A drop to 49\%, then all staff members go "westward"
immediately.

The above example reveals that the dispute in management idea is a kind of hidden
``contradiction" which is inevitable in a unified enterprise. But it does not exhibit
itself as two kinds of management action in opposite directions explicitly at the same
time. Notice further that the opinion itself of each side is also a ``contradiction",
because every reasonable thought must be a unity of opposites. This is one of the most
precious wisdom we have inherited from the philosophy for thousand years. Now
Eq.~(\ref{eq:1}) shows quite the similar thing. When $|\psi_1|>|\psi_0|$, to talk about
that $\psi_0$ describes a hidden clockwise current is incorrect in the sense of an
implicit (virtual) thing (like ``idea") being regarded as some explicit (real) thing
(like ``action"). In fact, a condition of $|\psi_0|>|\psi_1|$ is needed before $\psi_0$
can exhibit itself as a real clockwise current, but meanwhile, $\psi_1$ will then turn
from explicit into implicit at once.

The above comparison tells us that, just like the ``management idea" to some extent, the
wavefunction as shown by Eq.~(\ref{eq:1}) is by no means a visible substance but an
invisible ``contradiction field". This is why in quantum mechanics one should write down
a wavefunction, say a plane wave, in the following form:
\begin{eqnarray}
    \psi_p(x,t)&&=exp[i(px-Et)/\hbar]\nonumber\\
    &&=cos(kx-\omega t)+i sin(kx-\omega t)\label{eq:2},
\end{eqnarray}
where $p=\hbar k, E=\hbar\omega$. Here an imaginary number
unit $i=\surd(-1)$ is introduced such that its invisibility becomes obvious. Meanwhile,
two parts (both of them are real and represent the observable plane wave in classical
physics) separated by $i$ just represent two sides of the opposites in a contradiction
field.(see below, also [8]). If we simply express the wavefunction in Eq.~(\ref{eq:1}) by
$\psi_p$ in Eq.~(\ref{eq:2}), yielding:
\begin{equation}\psi=C_0\psi_p+C_1\psi_{-p}\label{eq:3},\end{equation}
Then we see immediately that when $C_0>C_1$, only clockwise current exists whereas only
counterclockwise current exists when $C_1>C_0$. Such kind of calculation can be performed
by everyone who had learnt quantum mechanics and we all know that the wavefunction is an
unobservable quantity. However, the puzzle is rooted at even deeper level.

Let us inquire of what is the meaning of $\psi_p$ in Eq.~(\ref{eq:2})? The explanation of
many text books in physics (including that written by me) is as follows. It describes the
motion of a free particle with energy $E$, momentum $p$ and $|\psi_p(x,t)|^2$ denotes the
probability of ``appearance of" the particle in point $x$ at time $t$.

Now I know I was wrong in the past. First, we had misunderstood the statistical
interpretation by Born in 1926. Now we have to correct the words ``appearance of" into
``measurement on". When we said ``appearance", we were tacitly assuming that the particle
is a ``point-particle" which moves to $x$ at time $t$ and then one might think
``naturally" that the Born statistical interpretation is equivalent to ``ensemble"
interpretation. Next, $x,p$ and $E$ are all ``information" about a particle and the
majority of physicists (including Einstein) considered that these information are
existing objectively, i.e., they exist already before the measurement is made. The
Copenhagen school led by Bohr and Heisenberg had different point of view, they emphasized
the repulsive property between the measurements of $x$ and $p$. Hence an uncertainty
relation emerges as;
\begin{equation}\triangle x\triangle p\geqslant\hbar/2\label{eq:4},\end{equation}
But they also cognized tacitly that the measurement is still a reflection process. So
they understood Eq.~(\ref{eq:4}) as a ``disturbance" of $p$ measurement on $x$
measurement or vice versa.

Now our point of view is the following: First, to think about a microscopic particle (say
an electron) as a point-particle is groundless in experiments. According to various
Schrodinger's cat experiments, the electron diffraction (double-slit) experiment and the
``which-way" experiment of atomic beam performed in 1998[9] (see also [3]), we prefer to
say that a particle has no fixed spatial extension and form. The electron in a hydrogen
atom is as big as the atom. And it could pass through the double-slit at a same time in
diffraction (see section~\ref{sec:5}). Second, a measurement is bound to change the
status of an object and there is no any information existing before the change of
object-status does occur. Hence the gain of information is not a process of reflection
but an outcome of the changing-process. In other words, the information is created by the
subject (via apparatuses) and the object in common.

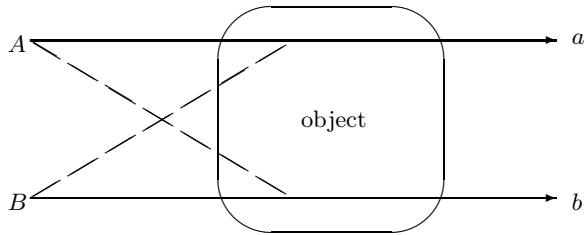
\begin{figure}[h]
\unitlength=1.0mm
\begin{picture}(80,30)
  \put(45,15){\oval(30,30)}
  \put(5,4.5){\vector(1,0){70.0}}
  \put(5,25.5){\vector(1,0){70.0}}
  \multiput(5,4.5)(5,3){7}{\line(5,3){4}}
  \multiput(5,25.5)(5,-3){7}{\line(5,-3){4}}
  \put(41,14){object}
  \put(2,3){$B$}
  \put(77,3){$b$}
  \put(2,24){$A$}
  \put(77,25){$a$}
\end{picture}
  \caption{\label{fig:1}
The measurement A(B) being an operation (denoted by the arrow) imposes on the object
(denoted by oval rectangle), creating the data a(b). If A is not in conformity with B,
then A(B) becomes the disturbance to b(a) as denoted by the dashed line.}
\end{figure}

The measurement is always an operation method (i.e., means) denoted by A for changing the
object and picking out corresponding data a: A$\rightarrow$a (see Fig.~\ref{fig:1}).
Similarly, another changing means B leads to b: B$\rightarrow$b. If A and B are not in
conformity but are imposed simultaneously on an object, then A (B) would become the
disturbance to b(a). Hence we see that Copenhagen school had confused the concept of
``changing" and that of ``disturbance". The changing is a necessity of getting the
information, but it shows up as a disturbance to other information at the same time. This
is the exact meaning of Eq.~(\ref{eq:4}), see [3].

Actually, different points of view are subjected to most stringent test so far in the
``which-way" experiment [9]. Since the which-way information is got from the internal
state instead of the impact of photon on the atom, the quantum coherence of atom's
center-of mass motion has not been destroyed. As no momentum transfer is measured, the
information about momentum $p$ itself does not exist at all, so does the relation (4). We
need not worry about something which has not emerged yet[3].

To further clarify the above point of view, we need a broad survey on philosophy.

\section{\label{sec:4}Philosophers' Way of Saying}
In ancient China, there was a saying in ``The Book of Rites-University" edited by
Confucius (551BC-479BC): To gain knowledge via 'gewu', the knowledge comes only after the
``wu"(object) is ``ge". The words ``gewu" was interpreted by the philosopher Cheng Yi
(1033-1107, in Song dynasty) as ``reaching the object" while Zhu Xi (1130-1200) explained
it as ``touching the object". In Ming dynasty, Wang Shou-ren (1472-1528) interpreted it
as ``seeing the object" and Wang Gen (1483-1540) explained it as ``the measurement on the
object", yielding a considerable progress (see [10]). It was not until Mao Zedong
(1893-1976) in his article ``on Practice", a principle of ``the cognition being stemming
from the changing" was stressed, implying that ``gewu" is now interpreted as some
``changing process" (``biange" in Chinese, for further explanation, see section
~\ref{sec:7}).It seems just appropriate. Just looking at the experimental methods in
modern physics evolving more and more abundant, the energy being raised higher and
higher, new phenomena and new particles emerge successively, we have been convinced that
the replacement of the ``reflection theory" by the ``changing theory" in the epistemology
of philosophy is indeed a big progress, a jump in conception.

In philosophy, one swam upstream from the epistemology to the ``ontology"---the inquiry
about the nature of universe and the origin of matters. It seems to us no surprise that
in conformity with the principle in epistemology that ``the information is generated from
the changing process", the ``noumenon" does not contain information. There were various
pronouns for the noumenon in Chinese philosophy, e.g., ``emptiness(void)" or ``oneness".
Sometimes, it was called the ``Tao" (which means the ``way" or ``law", see [10,11]). Lao
Tze (who lived in the same time with Confucius, maybe a little earlier) said: ``The Tao
that can be expressed is not the eternal Tao. The name that can be named is not the
permanent name". In our understanding, his saying implies that the fundamental (eternal
or perpetual) law cannot be expressed in words and the permanent name in wholeness (or
totality) cannot be divided and put into various categories. Actually, similar point of
view was prevailing in the Eastern philosophy, e.g., in the doctrine of Indian Hinduism
or Buddhism[11]. But it seems to us that a deep philosophy without explicit saying is
also a philosophy difficult to develope in real life. Lao Tze was wise to say more. He
said: ``the Tao generates one and one generates two...". Then after common efforts of
many philosophers, especially Wang Chong (27-100), Zhang Zai (1020-1077) and Wang Fuzhi
(1619-1692), the theory of ``yuanqi (the primary gas)" was developed. They claimed that
all matters are generated from ``yuanqi", inside which there are two opposites named
``yin and yang". It is the interaction and mutual transform between yin and yang that are
responsible for the motion and change of everything in the world. This is really a deep
and flexible ``ontology"[10,11].

In our opinion, to understand quantum mechanics at a deeper level, we have to deal with
``yuanqi" which was later called the ``Ether" in the Western philosophy or the ``vacuum"
in modern physics[3,8,11]. A particle is the excitation state of ``yuanqi", so its
wavefunction has two parts--the real and imaginary parts which are exactly the
mathematical expression of yin and yang as shown in Eq.~(\ref{eq:2}). Hence the
coordinate $x$ in the wavefunction must be the flowing coordinate of the ``field" rather
than that of the ``particle". After fixing $x$(or $t$) in Eq.~(\ref{eq:2}), we see the
growth and decline of yin and yang complementarily and periodically with the evolution of
$t$(or $x$). Notice further that the difference between yin and yang is merely relative.
At any time (or place) we can perform a phase (i.e., gauge) transformation:
$\psi\rightarrow exp(i\theta)\psi=\psi'$, and see that yin (or yang) transforming
immediately to its opposite yang (or yin), in which the property $i^2=-1$ plays a subtle
role. Moreover, as the counterpart of Eq.~(\ref{eq:2}) which describes a particle, the
wavefunction of its antiparticle is only different in the substitution of $i$ by
$(-i)$[8,3].

Now we turn to the Western philosophy. In the ancient Greek Philosophy, one school
represented by Herakleitos (~504BC) could be viewed as the counterpart of Chinese
philosophers like Lao Tze and Chuang Tze (369BC-286BC) [12]. But another two schools had
much more influence. One was the school represented by Pythagoras (~580BC), Plato (~428
BC) and Aristotle (384BC-322BC) who all emphasized numbers and mathematics. Another one
school represented by Demokritos (460BC-370BC) claimed the atomism. The philosophy of
these two schools had played an active role in promoting the development of modern
science. Therefore, quite naturally, physicists could make careless mistake to think
about the $x$ in a wavefunction as the spatial coordinate of some ``point-particle".

Beginning from Descartes (1596-1650), modern Western philosophers turned their emphasis
from the ``ontology" to the ``epistemology". They began to search for deeply the
relationship between ``being" and ``thinking", i.e., that between the ``object" and the
``subject". Here we wish to stress an important idea of Kant (1724-1804). He persisted in
separating so-called ``thing-in-itself" (or ``things-in-themselves") from
``phenomena"[13]. The former is existing objectively with no representation, i.e., we
(subject) have no knowledge of it. When the same thing exhibits itself as a phenomenon,
it turns to ``thing-for-us", i.e., a thing sensible to the subject. Kant's doctrine about
``thing-in-itself" was criticized by some philosophers and was often regarded as an
``idealist" or ``agnostic". But actually, the development of science has been proving
that Kant is right. As for the point that ``thing-in-itself contains no knowledge",
Kant's philosophy reached the same goal by different routes with the Eastern philosophy.

\section{\label{sec:5}Wavefunction is the Probability Amplitude of Fictitious Measurement}
It's not the time of Einstein and Bohr, many physicists don't pay enough attention to the
philosophy, they even think that philosophy has nothing to do with their researches.
However, the situation is quite the opposite. In a general way, everyone (let alone a
scientist) cannot detached from the philosophy. The problem is: Are you absorbing
nutritions from the whole treasure-house of humanbeing culture in a conscious manner? Or
you might be confined to some way of philosophical thinking while not be aware of. In the
specific sense of quantum mechanics, a series of experimental and theoretical researches
for over 70 years have been making unique contribution to the development of philosophy.
For instance, in 1935,the paper titled ``Can Quantum-Mechanical Description of Physical
Reality Be Considered Complete?" by Einstein, Podolsky and Rosen[14] not only initiated a
very important research field in physics called the EPR paradox[3], but also raised a
proposition in philosophy that ``what is the physical reality?"

Based on a series of experiments in recent decades, e.g., the two-photon EPR experiment,
physicists have been cognizing the strange entanglement phenomena in a system composed of
two or more particles. For instance, if Alice measured a photon being $x$-polarized, then
Bob at far apart must measured another photon being $y$-polarized at the same time. It is
a general feature of quantum nonlocality and denies definitely the so-called ``local
reality principle" proposed by Einstein. Besides, a beautiful EPR experiment on $K^0
\overline{K}^0$ system [15] together with its theoretical interpretation [16] verified
further that there must be antiparticle existing in quantum mechanics and its
wavefunction is merely different from that of particle in their opposite signs of phase.
Therefore, in our opinion, the EPR question as a ``paradox" is now basically over too.

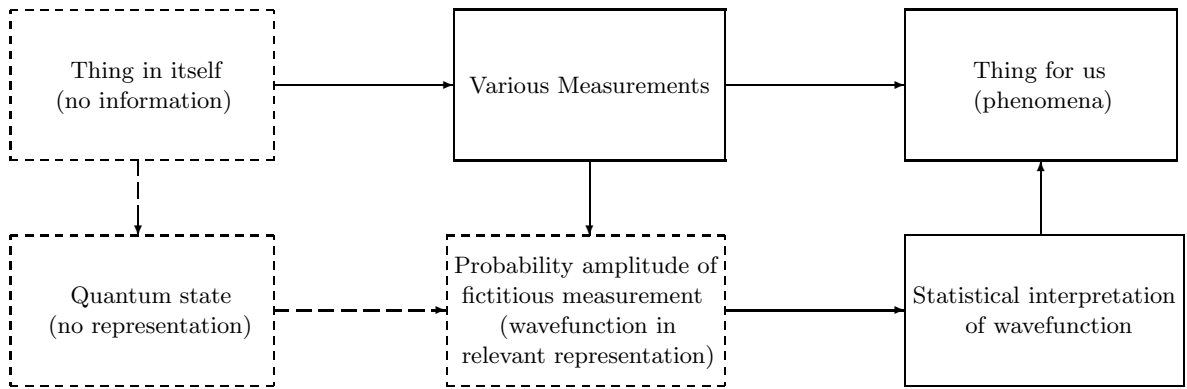
\begin{figure*}
\unitlength=1.0mm
\begin{picture}(170,55)
  \put(10,5){\dashbox(35,20){}}
  \put(18,16){Quantum state}
  \put(15,12){(no representation)}
  \put(68,5){\dashbox(37,20){}}
  \put(69,20){Probability amplitude of}
  \put(70,16){fictitious measurement}
  \put(75,12){(wavefunction in}
  \put(70,8){relevant representation)}
  \put(129,5){\framebox(37,20){}}
  \put(130,16){Statistical interpretation}
  \put(137,12){of wavefunction}
  \put(10,35){\dashbox(35,20){}}
  \put(18,46){Thing in itself}
  \put(16,42){(no information)}
  \put(69,35){\framebox(36,20){Various Measurements}}
  \put(129,35){\framebox(36,20){}}
  \put(138,46){Thing for us}
  \put(138,42){(phenomena)}
  \multiput(45,15)(3,0){7}{\line(1,0){2}}
  \put(66,15){\vector(1,0){2}}
  \put(105,15){\vector(1,0){24,0}}
  \put(45,45){\vector(1,0){24,0}}
  \put(105,45){\vector(1,0){24,0}}
  \multiput(27,35)(0,-3){2}{\line(0,-1){2}}
  \put(27,29){\vector(0,-1){4}}
  \put(87,35){\vector(0,-1){10}}
  \put(147,25){\vector(0,1){10}}
\end{picture}
  \caption{\label{fig:2}
The "reality" should be defined at two levels linked by the measurement. And quantum
mechanics works exactly in parallel. (Dashed lines are used to denote invisible things)}
\end{figure*}

I have been thinking about a diagram which could answer EPR's question from the viewpoint
combining both quantum mechanics and philosophy. After meditation for years, a diagram
came to me in an early morning as shown in Fig.~\ref{fig:2}.

In the upper part of Fig.~\ref{fig:2}, the epistemology and ontology are combined to
stress that it is the measurement, i.e., the process of changing, which realizes the
transform from ``thing-in-itself" to ``thing-for-us". The lower part shows how quantum
mechanics works. First, we divide from ``things-in-themselves" a system which is prepared
to be studied by means of boundary conditions and denote it by a quantum state (a
$|\psi>$ vector in the Hilbert space as introduced by Dirac). It has no representation
and so no coordinate $x$ ,even no time $t$ in the Heisenberg picture (exactly coinciding
with Kant's thinking). Next, in accordance with the phenomenon we are going to observe,
we choose an ``ideal apparatus" denoted by $|x,t>$ for measuring the quantity $x$ (say,
the position of a particle). Then a wavefunction in the corresponding $x$ representation
can be written down as:
\begin{equation}\label{eq:5}
\psi(x,t)=<x,t|\psi>
\end{equation}
which is the ``projection" (also called the contraction or scalar product) of a ``quantum
state vector" $|\psi>$ onto the ``coordinate basis vector" $|x,t>$.

We claim that a wavefunction is the probability amplitude of fictitious measurement with
the following explanation.

(a) This measurement is fictitious theoretically and not a realistic one. So it does not
destroy the coherence of quantum state. For a same quantum state, we may write down
several wavefunctions in different representations at the same time.

(b) A wavefunction in certain representation obtained via the fictitious measurement
reflects the existence of particle in the form of ``contradiction field" at each point of
the representation space. This kind of fictitious reflection is quite flexible. For
example, in nonrelativistic quantum mechanics, the wavefunction describing a stationary
motion of particle  displays precisely the contradiction equilibrium state reached via
interactions between the particle and its environment. In this case the energy E of the
particle contains merely the kinetic energy and potential energy. On the other hand, in
relativistic quantum mechanics, $E$ must includes the rest energy which becomes more
important. So the wavefunction will reflect the stronger (principal) contradiction field
inside the particle while still keep the meaning of contradiction field of the particle
with its environment[3].

(c) Let's discuss the double-slit diffraction of electrons as an example (for apparatus,
see [1,2,3]). Before an electron is detected by the detector on the screen at right-side,
its wavefunction at the right-side of double-slit (which is just the boundary condition
in present problem) is expressed as a linear superposition of two spherical waves
delivered from slit 1 and slit 2 respectively:
\begin{equation}\label{eq:6}
  \psi(x,t)=\psi_1(x,t)+\psi_2(x,t)
\end{equation}
In the expression:
\begin{equation}\label{eq:7}
  |\psi(x,t)|^2=|\psi_1(x,t)|^2+|\psi_2(x,t)|^2+2Re[\psi_1^*(x,t)\psi_2(x,t)].
\end{equation}
the third term at right-side is named the ``interference term" which is varying at
various points on the screen to be either positive or negative.

The electron beam density can be lowered to very weak level in experiments, say, there is
only one electron impacting at one point on the screen per second. Nonetheless, after a
long duration, interference fringes are displayed on the screen and can be interpreted by
Eq.~\ref{eq:7}. It was thought being too incredible because a proposition was raised in
the past that (see [1],p.139) ``either an electron goes through slit 1 or it goes through
slit 2" whereas the explanation of interference requires two waves coming from both slits
(as shown by the third term in Eq.~\ref{eq:7}). Hence a conceptual ``paradox" was
inevitable that ``the electron is a particle but also a wave at the same time".

As discussed by Feynman [1], an electron, before it is detected, can only be described by
the wavefunction, which spreads all over the space around the double-slit. Thus the above
proposition itself is meaningless. (We ask our readers to notice that instead of choosing
one answer from two possibilities destined by the proposition, we just negate the
proposition itself. Similar attitude will be adopted when other paradoxes in quantum
mechanics are dealt with, see [3]).

Once an electron is captured by an atom (in the detector), they will form a new quantum
correlated state. In other words, when an electron is detected at certain point $x$, it
exhibits itself as a real transform (transmutation) process of contradictions, which
renders the ``contradiction field" at other points (describing the fictitious interaction
between the electron and its environment) suddenly disappearing. It is the meaning of
so-called ``collapsing " of wavefunction toward the $x$ point.

Because of nonlocality (spatial extension) of an electron, also due to infinite degrees
of freedom in interactions between an electron and the screen (detector), they haven't
been and are impossible to be expressed in the Hamiltonian and boundary conditions, which
atom on the screen will ``seize" the electron is an event of probability. What the
wavefunction can tell us is a kind of potential possibility of contradiction transform at
point x. Just like other field in physics, the strength of contradiction field is
proportional to its amplitude. So $|\psi(x,t)|^2$ represents the real probability of
detecting the electron at point $x$. The above probability interpretation is different
from the ``ensemble interpretation" based on ``point-particle". The classical concept of
``ensemble" cannot reflect the subtlety of quantum probability unveiled in experiments on
single electron, ion or atom, let alone the strange entanglement in quantum system
compose of two or more particles (where the concrete correlation form is the outcome of
measurement). We appreciate very much the saying by Herbert[17]. He called the
wavefunction a ``possibility wave" and wrote down an interesting formula:
\begin{equation}\label{eq:8}
  probability=(possibility)^2.
\end{equation}
For further clarity, we suggest that one more word could be added to both sides
respectively as follows;
\begin{equation}\label{eq:9}
  |\psi|^2=real\;probability=(potential\;possibility)^2.
\end{equation}
We stress again that the reason why $\psi$ denotes the potential possibility is because
it, being a ``contradiction field", reflects the interaction between a ``fictitious
apparatus" and the object. Hence $\psi$ is something ``half-real but half-virtual" and is
located between the ``thing-in-itself" and the phenomenon, it is a reflection of the
existence state of an object in our cognition. The introduction of wavefunction is
undoubtly one of the greatest innovations in the human history, also an epoch-making
discovery in physics----a discovery that the essence of everything turns out to be
contradictions rather than some indivisible point-particles.

(d) Being a fictitious contradiction field, the wavefunction can be superposed linearly
and obeys the linear Schr$\stackrel{..}{\rm o}$dinger equation, showing that wave motion
is invisible and reversible. But once when a particle is measured , implying that a
realistic transform process occurs among contradictions, the particle property exhibits
itself immediately. The corresponding particle creation (and/or) annihilation operators
obey the Heisenberg (Fock) motion equation which is nonlinear in general, showing that
only during the real transform process can contradictions be visible---they cease to be
linearly superposed and become exclusive each other and irreversible. Hence, the
so-called ``wave-particle duality" is by no means a ``paradox" at the same level but in
conformity with the feature of particle being contradiction field.

(e) The concepts about position $(x)$, momentum $(p)$ and energy $(E)$ etc. which should
be derived originally from measurements via the analysis and induction method are now
introduced into the theoretical deduction process in advance by means of wavefunction
which is a contradiction field describing the interaction of fictitious apparatus with
the object. Then using the dynamical equation (Schr$\stackrel{..}{\rm o}$dinger equation,
Dirac equation, etc), one will be able to predict what observables with their
probabilities of occurrence will appear in a real measurement. This should be viewed as a
fusion and development at higher level of analysis---induction method and deduction
method which are originally separated in the history. The fusion of these two methods
makes quantum mechanics a scientific theory of contradictions which can be calculated
quantitatively, so it brings the human's subjective activity into full play in cognizing
the objective world.

\section{\label{sec:6}Mathematicians' way of saying}
To study quantum mechanics, one has to learn contemporary mathematics (topology, etc.).
But it was too difficult to me. Just like what described by Prof. C.N.Yang, facing a book
of this kind, I couldn't read further after reading one page. Sometimes, I even couldn't
read further after reading one line. Why formulas are much less than words in
mathematical books? Why there is no subscript under an English letter in these formulas?
I wondered for many years before I realized eventually that it is because mathematicians
pay more and more attention to separate the objective existence from its representation
rigorously.

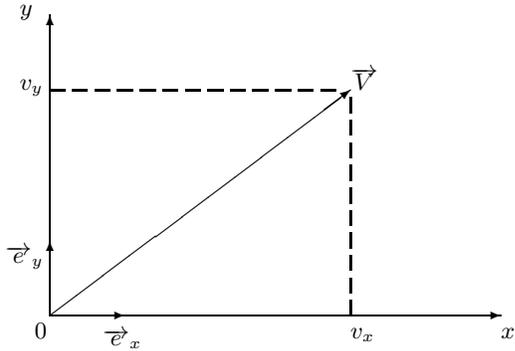
\begin{figure}
\unitlength=1.0mm
\begin{picture}(80,45)
  \put(10,5){\vector(0,1){10}}
  \put(10,5){\vector(1,0){10}}
  \put(10,15){\vector(0,1){30}}
  \put(20,5){\vector(1,0){50}}
  \multiput(50,5)(0,3){10}{\line(0,1){2}}
  \multiput(10,35)(3,0){13}{\line(1,0){2}}
  \put(10,5){\vector(4,3){40}}
  \put(50,35){$\overrightarrow{V}$}
  \put(8,2){0}
  \put(4,12){$\overrightarrow{e}_y$}
  \put(17,1){$\overrightarrow{e}_x$}
  \put(50,2){$v_x$}
  \put(70,2){$x$}
  \put(6,35){$v_y$}
  \put(6,45){$y$}
\end{picture}
  \caption{\label{fig:3}The concrete representation $v_x$ and $v_y$ of an abstract vector $V$ (in
two-dimensional space) are contractions (scalar product or projection) of $V$ with the
unit vectors of coordinate-system, $e_x$ and $e_y$.}
\end{figure}

Let's look at a familiar vector $V$, it merely denotes a ``geometrical object" having a
direction in the space. No any number is involved. To endow it with some concrete
representation, mathematicians introduce a coordinate-system (in two-dimensional space in
Fig.~\ref{fig:3} as an example) and project $V$ onto the coordinate axes, yielding $v_x$
and $v_y$. The latter are then the concrete representation of V. In rigorous notation,
they read:
\begin{equation}\label{eq:10}
  v_x=e_x\cdot V, \hspace{1.0cm} v_y=e_y\cdot V.
\end{equation}
where $e_x$ and $e_y$ are unit vectors along the $x$ and $y$ axes. We should notice the
resemblance between Eq.~(\ref{eq:10}) and Eq.~(\ref{eq:5}).

Furthermore, in contemporary mathematics, a vector field continuously spreading over the
space, such as the vector potential field $A(x,t)$ of electromagnetic field in physics,
is needed to be discussed. Because not only the components of $A(A_x,A_y)$ are depending
on the coordinate-system, but $A$ is a function of $x$ as well, mathematicians innovate a
very elegant notation like $A$ to express that:
\begin{equation}
  A=A_\mu dx_\mu=A_xdx+A_ydy=A'_xdx'+A'_ydy'.
\label{eq:11}\end{equation}

Here $A$ is expanded by its components while the coordinate differentials $dx_\mu$ are
written together. Such kind of form is called the ``connection one-form". It is actually
the counterpart or development of the notation for single vector $V$ as :
\begin{equation}\label{eq:12}
   V=v_{x}e_x+v_{y}e_y=v'_xe'_x+v'_ye'_y.
\end{equation}

The common feature of Eqs.~(\ref{eq:11}) and ~(\ref{eq:12}) is that the ``geometrical
object" at the left-side is linked to its concrete representation via the
coordinate-system. Thus a fact is stressed that the existence of a geometrical object is
independent of the subjective choice among various coordinate-systems whereas its
representations are taken in different concrete forms with respect to different
coordinate-systems.

I would like to tell a small story I heard it myself. Once a time, the famous
mathematician, Prof. S-S Chern delivered a lecture on contemporary differential geometry
in Nankai University. When he wrote down a letter $A$, among those present was Prof.
C.N.Yang who suddenly asked: ``There is no subscript in this $A$. Did you write it in
this way when you were doing research?" Then Prof. Chern replied immediately: ``No. I
used to keep all subscripts during my calculation but threw them away when the
calculation was done (and writing my paper)."

The conversation between Profs. Yang and Chern revealed that actually mathematicians are
not so different from physicists when they think about some problem and begin to do the
research. They nearly all begin ``from the particular to the general". However, once they
write the book, their styles are widely different. It reflects a deep discrepancy between
the ``culture" of mathematics and that of physics. This discrepancy is often difficult to
understand if not listening to the lecture personally and exchanging the point of view
thoroughly. (I wish to thank Prof. Yi-zhi Huang for discussing this problem with me via
the phone-call many times.)

\section{\label{sec:7}Principle of Relativity and Principle of Changing in Epistemology}

All our discussions above are based on a ``principle of relativity" in epistemology: When
Einstein established the theory of special relativity, he taught us that one should not
discuss the absolute motion, absolute space or absolute time. Rather, one should discuss
the relative motion, relative space or relative time. In general, when one is talking
about cognition, one must first put himself into the process of cognition as the
``subject" and then survey the environment of the object under consideration. A thing can
only be cognized during its motion and change relative to other things. If isolated from
its opposites (i.e., its environment, including the subject and all measuring
apparatuses), it merely exists abstractly and so is bound to become mysterious object
devoid of any understanding.

In the Chinese classical novel titled ``Dream of the Red Mansion" (by Cao Xue-qin, who
told a story occurred in a big Jia's family in Beijing during 1729-1737 in Qing dynasty),
a young man Jia Baoyu was worried about Ms. Lin Daiyu who used to feel sad all day long.
``I can't understand her", Baoyu thought. Until one day he suddenly got an inspiration
and wrote down as follows:

You won't be you if I wasn't born.

One can't understand her from her alone.

Therefore, to understand an abstract quantum state $|\psi>$ of a particle, one brings his
subjective activity into full play by introducing a fictitious apparatus $|x,t>$,
yielding the wavefunction $\psi(x,t)=<x,t|\psi>$ to reflect the ``contradiction" due to
the interaction between the particle and the fictitious apparatus for measuring $x$ ($x$
is not necessary a continuous spatial coordinate). By calculation of $\psi$, one can
predict in advance what are potential observables and what are their probabilities of
occurance before the experiment is really made.

Similarly, to understand Daiyu, one has to survey her environment. Being a sensitive
young lady, Daiyu was living in the global environment of Jia's big family. If isolated
from her relations with Baoyu, Baoyu's grandmother, Ms. Wang Xi-feng and Ms. Xue Bao-cha,
etc., i.e., if isolated from all realistic appearances of these relations and their
contradiction analysis, Daiyu would be never understood why she was so sad.

After reading what Baoyu wrote, Daiyu, with Ms. Xue and Ms. Shi Xiangyun, went
immediately to visit Baoyu for a conversation, aiming at learning what he was thinking at
that time. But was it a simple "thought exchange or reflection?" No. Just at the instant
their conversation began, everyone of them was changed onceagain and Daiyu was caught by
a new suffering.

Hence, a measurement is by no means a process of ``reflection", but a process of changing
(``biange") the object by the subject. However, the sole English word ``changing" is far
from enough to express the deep connotation of Chinese words ``biange", which implies
``penetrating deeply into the object and changing it at the root". The root is nothing
but ``contradictions". So the changing process is actually a process of transform
(transmutation) between contradictions. In terms of newest terminology in quantum
mechanics, during the measurement, a new quantum correlated state (entangled state) is
established between the apparatus and object. Meanwhile, the quantum (coherently)
correlated state originally existing inside the object is destroyed. It is just during
the process of destruction and reestablishment of the equilibrium state among
contradictions, can some ``information" be created and is sensible to the subject.

\section{\label{sec:8}Epilogue}
I learnt quantum mechanics in 1954 when Prof. Xie Xide was our teacher. Ever since, I
have been involved in a strong feeling with quantum mechanics. I was so keen on it, so
excited in it and sometimes even nervous  as well. It was such a feeling just like being
haunted by a cat in the dark. It was not until recent years that  a series of new
brilliant experiments showed up. The light suddenly flashed in front of us. I was
pleasantly surprised and began to relax for a while. Then a poem came to my mind as
follows:

Now the cat looks amazing.

Let's listen to what she's saying:

``People were surprised at the emperor's clothes.

But he was wearing nothing."

\end{document}